# Dispersion-Engineered Terahertz Silicon Interconnects Enabling Terabit-Scale Data Links

*Bodhan Chakraborty[1,2], Wenhao Wang[3], Nikhil Navaratna[1,2], Thomas Caiwei Tan[1,2], Pascal Szriftgiser[4], Hadjer Nihel Khelil[5], Guillaume Ducournau[5], and Ranjan Singh[6]**

[1] Division of Physics and Applied Physics, School of Physical and Mathematical Sciences, Nanyang Technological University, Singapore, 637371, Singapore

[2] Centre for Disruptive Photonic Technologies, The Photonics Institute, Nanyang Technological University, Singapore, 639798, Singapore

[3] Department of Electronic and Information Engineering, School of Engineering, Westlake University, Hangzhou 310030, China

[4] Université de Lille, CNRS, UMR 8523 – PhLAM, Laboratoire de Physique des Lasers, Atomes et Molécules, Lille, France

[5] Université de Lille, CNRS, UMR 8520 – IEMN, Institut d′Electronique Microelectronique et Nanotechnologie, Lille, France

[6] Department of Electrical Engineering, University of Notre Dame, Notre Dame, IN 46556, USA

* Corresponding author: rsingh3@nd.edu

**Abstract:**

The rapid growth of artificial intelligence (AI) and data-centric computing demands exabyte-scale data transfer, pushing data centres and high-performance computing toward fundamental limits where interconnects increasingly dominate performance and energy consumption. Optical interconnects enable high-capacity, long-reach communication, but their complexity and energy overhead constrain scalability in the short-reach regimes central to chiplet-based and on-chip systems. Terahertz (THz) silicon interconnects offer a promising solution for centimetre-scale on-chip and chip-to-chip communication by bridging electronics and photonics, enabling high bandwidth and low latency in simplified, complementary metal-oxide-semiconductor (CMOS)-compatible architectures. However, practical realization is challenged by stringent requirements for multi-band transmission, dual-polarization support, low-loss propagation, low group-velocity dispersion (GVD), and terabit-per-second throughput, with Bragg-induced stopbands and dispersion penalties limiting performance at high THz frequencies. Here, we overcome these constraints and demonstrate a CMOS-compatible, centimetre-scale, multi-band on-chip THz data link achieving 1.004 Tbps aggregate throughput. This record rate is enabled by suppressing Bragg-induced stopbands using dispersion-engineered, effective-medium-supported, unclad silicon waveguides, yielding flat transmission and low-ripple group delay across multiple THz bands. The waveguide platform operates from 220 to 500 GHz, supporting transverse-electric (TE) and transverse-magnetic (TM) polarizations with low path loss, minimal bending loss, and low GVD. Fourteen channels in a straight waveguide and twelve channels in a 90° bend achieve aggregate data rates of 1.004 Tbps and 0.895 Tbps, respectively, with GVD as low as 0.15 $ps^2/mm$ over the full band. These results establish a scalable, energy-efficient platform for multi-band THz interconnects for high-density on-chip and chip-to-chip communication fabrics that address the bandwidth and energy bottlenecks of next-generation AI systems and emerging 6G technologies.

## Introduction

The rapid growth of internet services and artificial intelligence (AI) has driven data traffic to exabyte (EB) scales, pushing data-centre (DC) and high-performance computing (HPC) infrastructures to their limits[1]. By 2030, ~70% of global DC capacity is expected to be devoted to AI workloads, while wide-area network (WAN) traffic is projected to reach 3,386 EB per month by 2033, including 1088 EB from AI, corresponding to a compound annual growth rate (CAGR) of 24%[2]. Optical interconnects enable high-capacity, low-latency data movement across DC and HPC infrastructures. However, their reliance on complex electro-optic transceivers results in substantial energy overheads and limits scalability for short-reach, chip-scale links[3]. The terahertz (THz) band (0.1-10 THz) offers vast spectral resources, high carrier frequencies, and low latency, providing a pathway to terabit-per-second (Tbps) data transmission[4–9]. THz interconnects are particularly promising for chiplet fabrics, where centimetre-scale links with nanosecond latency could bridge the performance gap between on-chip electronics and off-chip optics while remaining complementary metal-oxide-semiconductor (CMOS)-compatible[10]. Conventional metallic THz interconnects are fundamentally constrained by their low bandwidth-distance product, limiting their suitability for such applications[11,12]. By contrast, THz silicon waveguide-based interconnects fabricated from high-resistivity float-zone intrinsic silicon (HR-Si) offer a compelling alternative[12]. These structures enable low-loss THz propagation, strong mode confinement due to the high refractive index contrast ($n_{\mathrm{Si}}$ = 3.42), and compatibility with CMOS fabrication through relatively simple single-mask etching processes[13].

Despite these advantages, a primary challenge for THz silicon waveguide-based interconnects is achieving CMOS-compatibility, centimetre-scale, multi-band transmission with low loss and low group-velocity dispersion (GVD) while supporting multi-carrier operation[6,14,15]. A range of THz silicon waveguide platforms have been demonstrated, including photonic crystal waveguides[16–18], topological valley photonic crystal waveguides[19–21], effective-medium-clad waveguides[14,22,23], and unclad waveguides[24–26]. These have enabled the integration of passive components such as power splitters[25] and multi/demultiplexers[21,23,24], as well as on-chip active devices including sources[27], detectors[17], and mixers[28]. However, most platforms remain limited in operational bandwidth, often confined to a single THz band[13]. Approaches to extend bandwidth across multiple THz bands have therefore relied on microscale supports that preserve high index contrast while accommodating both transverse-electric (TE) and transverse-magnetic (TM) polarizations. For example, effective-medium-clad waveguides employ deep sub-wavelength through-hole arrays etched into HR-Si, analogous to photonic crystals[14,23,29], but their fabrication requires high-aspect-ratio features that challenge scalability. Suspended silicon beams maximize index contrast by eliminating in-plane cladding, yet suffer from mechanical fragility[30]. Unclad waveguides offer a compromise by confining the effective-medium claddings to the device terminations, thereby suppressing leakage while preserving index contrast along the propagation axis[25,31]. However, a fundamental limitation across these THz silicon interconnects is Bragg scattering, which occurs when structural periodicity approaches the guided wavelength ($\lambda_g$) in the silicon core, producing transmission stopbands that severely constrain multi-band operation[32,33].

Here, we report the first demonstration of on-chip terabit-per-second THz communication (Table 1), enabled by THz interconnects fabricated on HR-Si and based on a dispersion-engineered, effective-medium-supported, unclad silicon waveguide design that suppresses Bragg scattering at high frequencies. The CMOS-compatible, centimetre-scale platform supports multi-band (multi-carrier) operation across two adjacent THz bands (WR-3.4 and WR-2.2), spanning 220-500 GHz, while accommodating both transverse-electric (TE) and transverse-magnetic (TM) polarizations and maintaining low path loss and low GVD. We systematically quantify the average insertion, path, and bending losses using straight waveguides with varying transmission link lengths, together with waveguides incorporating 90° and 180° bends with subwavelength bending radii ($R \approx 0.83\lambda$ at 500 GHz), which preserve identical link distances while differing in the number of turns. The average insertion loss associated with coupling between the hollow metallic waveguides and the integrated tapered couplers used for chip-level THz coupling is 0.21 dB and 0.48 dB for TE and TM polarizations, respectively. The corresponding average path loss is 0.22 dB/cm (TE) and 0.19 dB/cm (TM). The average bending losses are 0.35 dB (TE) and 0.37 dB (TM) per turn for 90° bends, and 0.39 dB (TE) and 0.56 dB (TM) per turn for 180° bends, evaluated over the 220-500 GHz band. Using a combined photonic transmitter (Tx) and electronic receiver (Rx) architecture, we demonstrate aggregate on-chip data rates of 1.004 Tbps and 0.895 Tbps across the 220-500 GHz band using TE-polarized THz waves in straight waveguides and waveguides incorporating 90° bends, respectively. These performances are enabled by managing the dispersion, yielding GVD values as low as 0.18 and 0.15 $ps^2$/mm, extracted on a per-channel, frequency-domain basis across multiple THz channels spanning the WR-3.4 and WR-2.2 bands. The straight and bent waveguides support 14 and 12 channels with distinct carrier frequencies, respectively, employing high-order complex modulation formats. To our knowledge, no prior work has demonstrated THz waveguides that suppress Bragg stopbands while simultaneously supporting multiple THz bands with low loss, low GVD, and near- or terabit-per-second data transmission using multiple THz carriers in both straight waveguides and bends. Together, these results establish dispersion-engineered, effective-medium-supported, unclad silicon waveguides as a viable platform for multi-band, low-loss THz interconnects, and open new pathways toward integrated electronic-photonic THz systems with higher per-waveguide data rates, increased channel counts, and enhanced spectral efficiency.

**Table 1 |** Comparison with state-of-the-art THz silicon waveguide technologies on an HR-Si platform.

| Interconnect Technology | 3-dB Bandwidth (GHz) | Tx (Rx) | Link Distance (cm) | Complex Modulation Format | No. of Channels | Data Rate | | Min. EVM[c] (%) |
|---|---|---|---|---|---|---|---|---|
| | | | | | | Max[a] (Gbps) | Aggregated[b] (Tbps) | |
| ***Single Channel*** | | | | | | | | |
| EM WG (On-chip)[22] | 130 | UTC-PD (RTD) | 5 | QAM-32 | 1 | 100 | 0.100 | N.A. |
| VPC WG[19] | 30 | UTC-PD (SHM) | 1 | QAM-32 | 1 | 160 | 0.160 | 8.2 |
| VPC WG[20] | 29.6 | UTC-PD (SHM) | 1 | QAM-16 | 1 | 108 | 0.108 | N.A. |
| ***Multiple Channels*** | | | | | | | | |
| PC WG[16] | 90 | UTC-PD (SHM) | 2.49 | QAM-32 | 4 | 100 | 0.327 | 5 |
| EM WG[23] | 105 | UTC-PD (FMBD) | 2 | QAM-16 | 2[d] | 80<br>100 | 0.155<br>0.190 | 12.69 |
| UC WG[24] | 90 | N.A. | 2.26 | QAM-16 | 4 | 50 | 0.200 | 5.6 |
| VPC WG[21] | 24.5 | UTC-PD (SHM) | 2.17 | QAM-64 | 2 | 75 | 0.150 | 5 |
| **This work** | **280**[e]<br>**192.39**[f] | **UTC-PD (SHM)** | **2.202** | **QAM-16, 32, 64 and APSK-16** | **14**<br>**12** | **90** | **1.004**<br>**0.895** | **4.24**<br>**4.12** |

Text in bold – Our work in this article

PC: photonic crystal; EM: effective medium; VPC: valley photonic crystal; UC: unclad; WG: waveguide; Tx: transmitter; Rx: receiver; QAM: quadrature amplitude modulation; APSK: amplitude-phase-shift keying; EVM: error vector magnitude; UTC-PD: uni-travelling carrier photodiode; RTD: resonant tunneling diode; SHM: sub-harmonic mixer; FMBD: Fermi-level managed barrier diode; N.A.: not available.

[a] Maximum single-channel data rate.

[b] Total aggregate data rate obtained by summing contributions from all channels.

[c] Minimum error vector magnitude (EVM) measured in the THz communication experiments.

[d] Polarization multiplexing implemented within a single channel.

[e] Measured 3-dB bandwidth of the straight waveguide (WB1) device for TE polarization.

[f] Measured 3-dB bandwidth of the 90° bent waveguide (90B1) device for TE polarization.

## Results and Discussion

The all-silicon substrateless waveguide is fabricated on high-resistivity float-zone silicon (HR-Si, resistivity > 20 kΩ.cm; see Methods) and is based on a dispersion-engineered, effective-medium-supported, unclad waveguide platform. Fig. 1a shows an optical image of the straight waveguide device (WB1), with dimensions of 1.48 cm × 0.74 cm. The silicon core, with width $w$ = 0.26 mm, is supported at both ends by effective-medium claddings (0.29 cm × 0.34 cm), formed by a periodic array of circular through-holes with diameter $d$ = 110 μm and lattice constant $a$ = 120 μm. A 0.91 cm-long central unclad section maximizes the refractive-index contrast and leverages total-internal-reflection-based guidance to reduce scattering, yielding lower-loss and weakly dispersive propagation. Two 2 mm-wide silicon slabs provide additional mechanical stability during handling and characterization. Integrated tapered couplers ($P_1$ and $P_2$) at both ends enable efficient interfacing with hollow metallic waveguides. The THz signal (polarization indicated by blue double-sided arrows) is coupled into the waveguide through $P_1$, propagates along the silicon core (indicated by single-sided red arrows), and is coupled out through $P_2$, supporting operation across the WR-3.4 (220-330 GHz) and WR-2.2 (330-500 GHz) bands. The platform supports both transverse-electric (TE) and transverse-magnetic (TM) polarizations, with fundamental modes excited owing to strong mode matching at the coupling interface with the hollow metallic waveguides (see Supplementary Information S2).

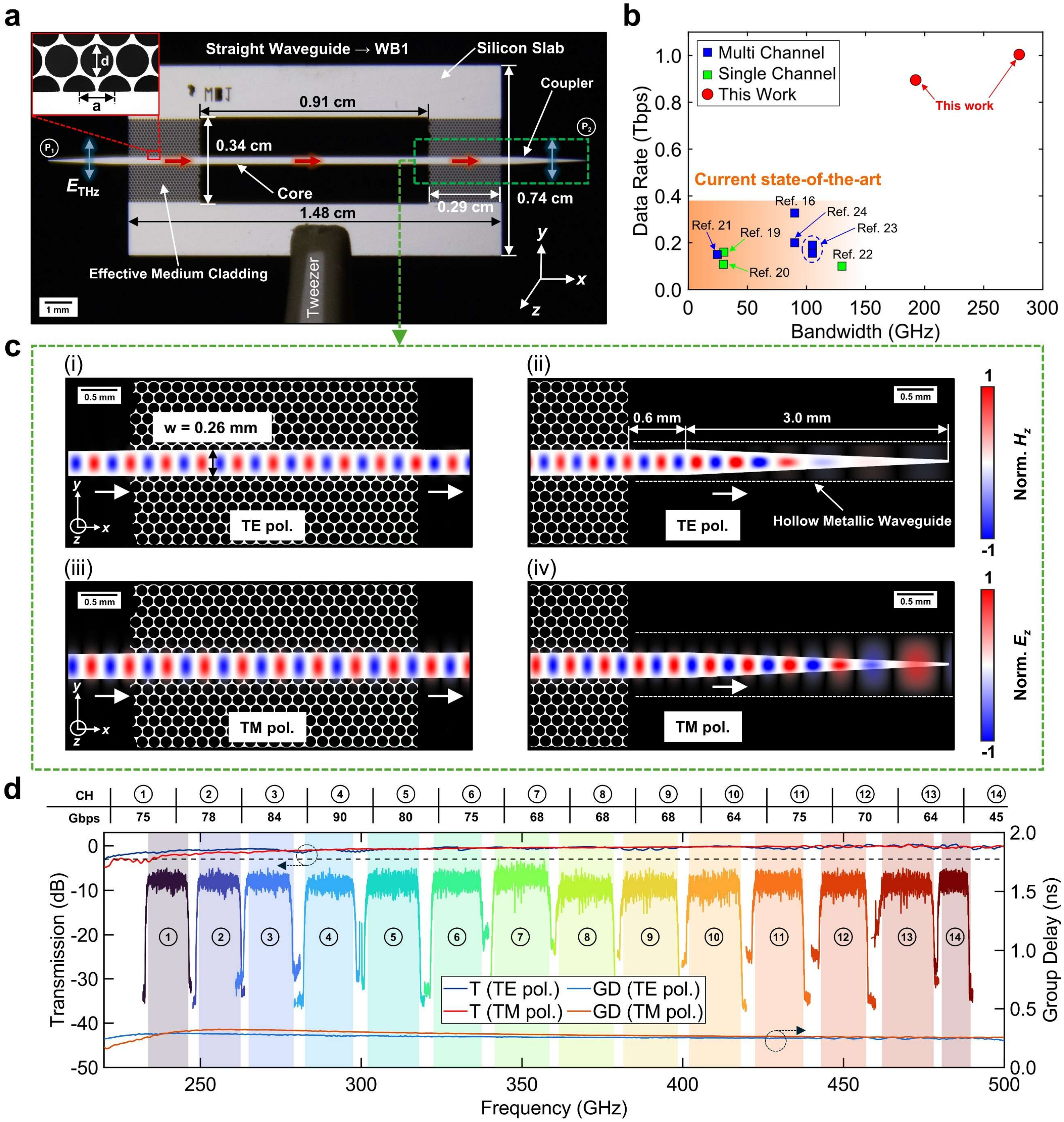

**Fig. 1 | THz silicon interconnects for terabit-per-second on-chip data communication. a**, Optical image of an all-silicon waveguide sample (WB1) with device dimensions of 1.48 cm × 0.74 cm. THz signals (polarization indicated by double-headed blue arrows) are coupled at both ends via integrated tapered couplers ($P_1$ and $P_2$) of length 3.6 mm and propagate along the silicon core (width $w = 0.26$ mm), as indicated by the red arrows. The inset shows the periodicity ($a = 120$ µm) and hole diameter ($d = 110$ µm) of the etched through-holes forming the effective-medium cladding that surrounds and supports the waveguiding core. A 2-mm-wide silicon slab is incorporated for mechanical support, and tweezers are used to hold the sample for alignment and measurement. The unclad region measures 0.91 cm × 0.34 cm, while the effective-medium cladding supporting the silicon core measures 0.29 cm × 0.34 cm. **b**, Comparison of the data rates achieved using the straight waveguide (WB1) and the 90° bent waveguide

(90B1) with previously reported on-chip THz data communication systems employing single- and multi-channel architectures based on THz silicon waveguides technologies on an HR-Si platform[16,19–24], highlighting the record data transfer rates of 1.004 Tbps and 0.895 Tbps for the WB1 and 90B1 devices, respectively. **c**, Microscope images of the core and tapered coupler regions of the WB1 sample, together with simulated field distributions for TE ($H_z$) and TM ($E_z$) polarizations at 330 GHz, showing subwavelength (sub-$\lambda_g$) field confinement in the silicon core. The normalized $z$-component of the magnetic field (Norm. $H_z$; TE polarization) is shown in (i) and (ii), and the normalized $z$-component of the electric field (Norm. $E_z$; TM polarization) in (iii) and (iv), for the core and tapered coupler regions (highlighted by green dashed lines in (**a**)), respectively. The hollow metallic waveguides used to couple the THz signals in and out are indicated by white dashed lines. **d**, Measured transmission ($S_{21}$; T) and group delay (GD) of the WB1 sample for both TE and TM polarizations. Fourteen channels are excited non-simultaneously, indicated by the shaded regions, each employing a different complex modulation format. The corresponding data rates, channel bandwidths, and normalized channel power spectra (offset by -5 dB for clarity) are shown for each channel.

Undesired stopbands at higher THz frequencies limit the bandwidth in effective-medium-cladding-based designs[33]. These stopbands arise when the hole periodicity approaches the guided wavelength ($\lambda_g$) in the silicon core, causing effective-index ($n_{eff}$) modulation[32]. We overcome this limitation and further enhance the bandwidth by modifying the through-holes adjacent to the waveguiding core from circular to semi-circular geometries, thereby suppressing Bragg scattering, flattening transmission ($S_{21}$) spectra, and minimizing group-velocity dispersion (GVD) to enable multiband operation. Microscope images of the waveguide core and tapered coupler (0.26 mm × 0.6 mm rectangular section followed by a 3 mm linear taper) inserted into the hollow metallic waveguide (white dashed lines), as shown in Fig. 3c, ensure efficient coupling of guided THz energy. Simulated normalized $z$-components of the magnetic field (Norm. $H_z$; Fig. 1c(i, ii)) and electric field (Norm. $E_z$; Fig. 1c(iii, iv)) for TE and TM polarizations, respectively, at 330 GHz are also shown. The field distributions reveal strong subwavelength (sub-$\lambda_g$) confinement of both electric and magnetic fields within the silicon core. Measured transmission ($S_{21}$) and group-delay characteristics of WB1 (Fig. 1d; see Methods for details of the experimental setup) demonstrate broadband, low-ripple propagation from 220 to 500 GHz, corresponding to a 280 GHz 3-dB bandwidth for both polarizations across multiple THz bands. To our knowledge, this represents the largest reported 3-dB bandwidth for THz silicon waveguides. Fourteen non-simultaneously excited channels span this bandwidth, employing complex modulation formats including QAM-16, 32, and 64, and APSK-16. The extracted channel bandwidths, normalized channel power spectra (offset by -5 dB for clarity), and associated data rates across multiple THz bands are shown in Fig. 1d. Benchmarking against previously reported single- and multichannel on-chip THz communication systems (Fig. 1b) highlights record aggregate data rates of 1.004 Tbps in the straight device (WB1) and 0.895 Tbps in the 90° bent counterpart (90B1), enabled by the substantially broadened 3-dB bandwidth.

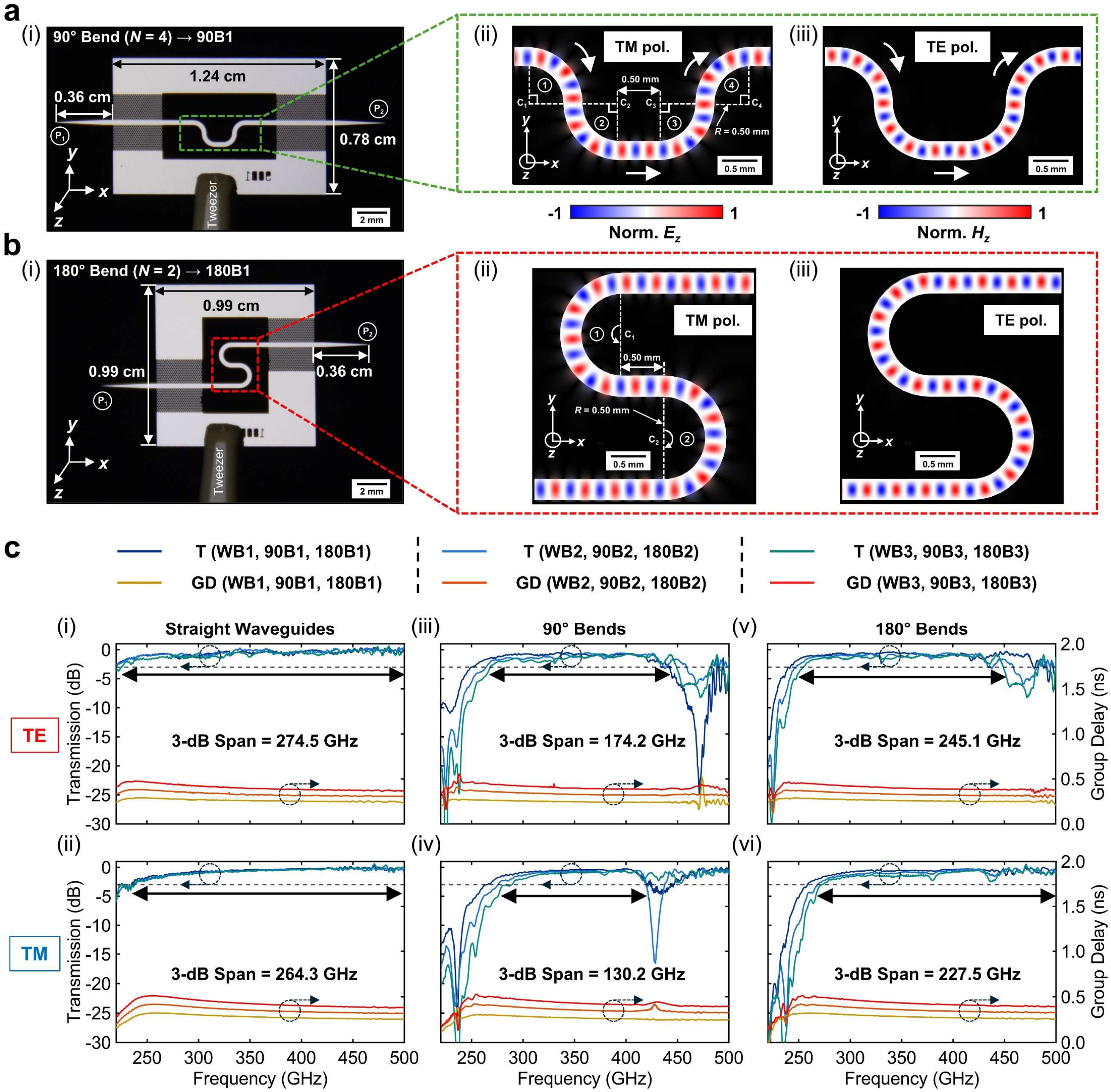

**Fig. 2 | Characterization of transmission and group delay in all-silicon straight, 90° bent, and 180° bent interconnects with varying transmission link lengths and number of turns per sample (*N*).** (**a**-i) and (**b**-i) show optical images of the 90° and 180° bent waveguide samples with $N = 4$ and $N = 2$, respectively. The chip dimensions are 1.24 cm × 0.78 cm for the 90° bent waveguide and 0.99 cm × 0.99 cm for the 180° bent waveguide. For both samples, a 0.36-mm-long integrated tapered coupler is used to couple the THz signal in and out through ports $P_1$ and $P_2$, as indicated in (**a**-i) and (**b**-i). The silicon core width of the bent waveguide samples is $w = 0.26$ mm. (**a**-ii) and (**a**-iii) show the microscope image of the 90° bend (90B1 sample) together with the normalized $z$-component electric (Norm. $E_z$) and magnetic field (Norm. $H_z$) distributions for TM and TE polarizations, respectively, in the region highlighted by green dashed lines in (**a**-i). Similarly, (**b**-ii) and (**b**-iii) show the microscope image of the 180° bend (180B1 sample) along with the Norm. $E_z$ and Norm. $H_z$ distributions for TM and TE polarizations, respectively, in the region highlighted by red dashed lines in (**b**-i). All field distributions are computed at the centre frequency of 330 GHz. Also indicated in (**a**-ii), (**a**-iii), (**b**-ii), and (**b**-iii) are the subwavelength

bending radius ($R$) of 0.50 mm ($R \approx 0.83\lambda$ at 500 GHz), the 0.50-mm-long straight silicon core segment between successive turns, and the centre of each turn, labelled as $C_i$. (**c**-i, ii), (**c**-iii, iv), and (**c**-v, vi) show the transmission ($S_{21}$) and group delay of the straight waveguides (WB1, WB2, and WB3), 90° bent waveguides (90B1, 90B2, and 90B3), and 180° bent waveguides (180B1, 180B2, and 180B3), respectively, for both TE and TM polarizations. The 3-dB bandwidth regions are also highlighted in (**c**-i–vi).

Based on our design strategy, we fabricated multiple straight waveguide samples with increasing transmission link lengths of $L$ = 2.202 cm (WB1), 2.698 cm (WB2), and 3.194 cm (WB3). To address THz signal routing for dense on-chip integration without compromising bandwidth or signal fidelity, we further implemented two classes of bends (90° and 180°) derived from the straight-waveguide geometry, each incorporating different number of turns corresponding to the bend angle. For the 90° designs, the number of turns is $N$ = 4 (90B1), 8 (90B2) and 12 (90B3). Similarly, for the 180° designs, $N$ = 2 (180B1), 4 (180B2) and 6 (180B3). For each configuration, the total transmission link length is kept identical to that of its straight counterpart. For example, a straight waveguide (WB1), a 90° bent waveguide (90B1), and a 180° bent waveguide (180B1) all feature the same link length. Figure 2 presents the transmission and dispersion characteristics of these straight and bent THz silicon waveguides. Optical images of representative 90° (90B1) and 180° (180B1) devices are shown in Fig. 2a(i) and 2b(i), revealing compact chip dimensions of 1.24 cm × 0.78 cm and 0.99 cm × 0.99 cm, respectively, enabled by tight subwavelength bends with a radius $R$ = 0.50 mm ($0.83\lambda$ at 500 GHz). A 0.50 mm straight segment is inserted between successive turns to mitigate mode mismatch and suppress scattering of the fundamental mode into higher-order modes. In both geometries, integrated tapered couplers (marked as $P_1$ and $P_2$) ensure efficient THz in- and out-coupling. Simultaneously, the silicon core width is maintained at 0.26 mm to preserve consistent modal confinement across the THz bands. Microscope images of the bend regions, together with simulated normalized $z$-components of the electric and magnetic field (Fig. 2a – Fig. 2b) for TM and TE polarizations, respectively, at 330 GHz, confirm confined low-loss propagation through the 90° and 180° turns, with minimal perturbation of the guided mode. The field profiles indicate low radiation loss and preserved polarization integrity across successive turns. Figures 2c(i–vi) compare the measured transmission ($S_{21}$) and group delay of the straight waveguides (WB1–WB3), 90° bent waveguides (90B1–90B3), and 180° bent waveguides (180B1–180B3) for both TE and TM polarizations. All devices maintain flat transmission and low-ripple group delay throughout their 3-dB bandwidth, indicating low-loss and low-dispersion signal propagation. The transmission dips for the bent waveguides at frequencies above 400 GHz can be attributed to higher-order mode scattering (see Supplementary Information S3). At frequencies below 300 GHz, weaker confinement of the fundamental mode within the silicon core in the bent waveguides leads to reduced transmission compared with their straight-waveguide counterparts. In this regime, radiation losses at the bends dominate because longer wavelengths produce larger mode sizes relative to the subwavelength bend radius. Consequently, the effective index contrast decreases, and the critical angle for total internal reflection is exceeded at the bend, leading to increased radiation into the surrounding medium.

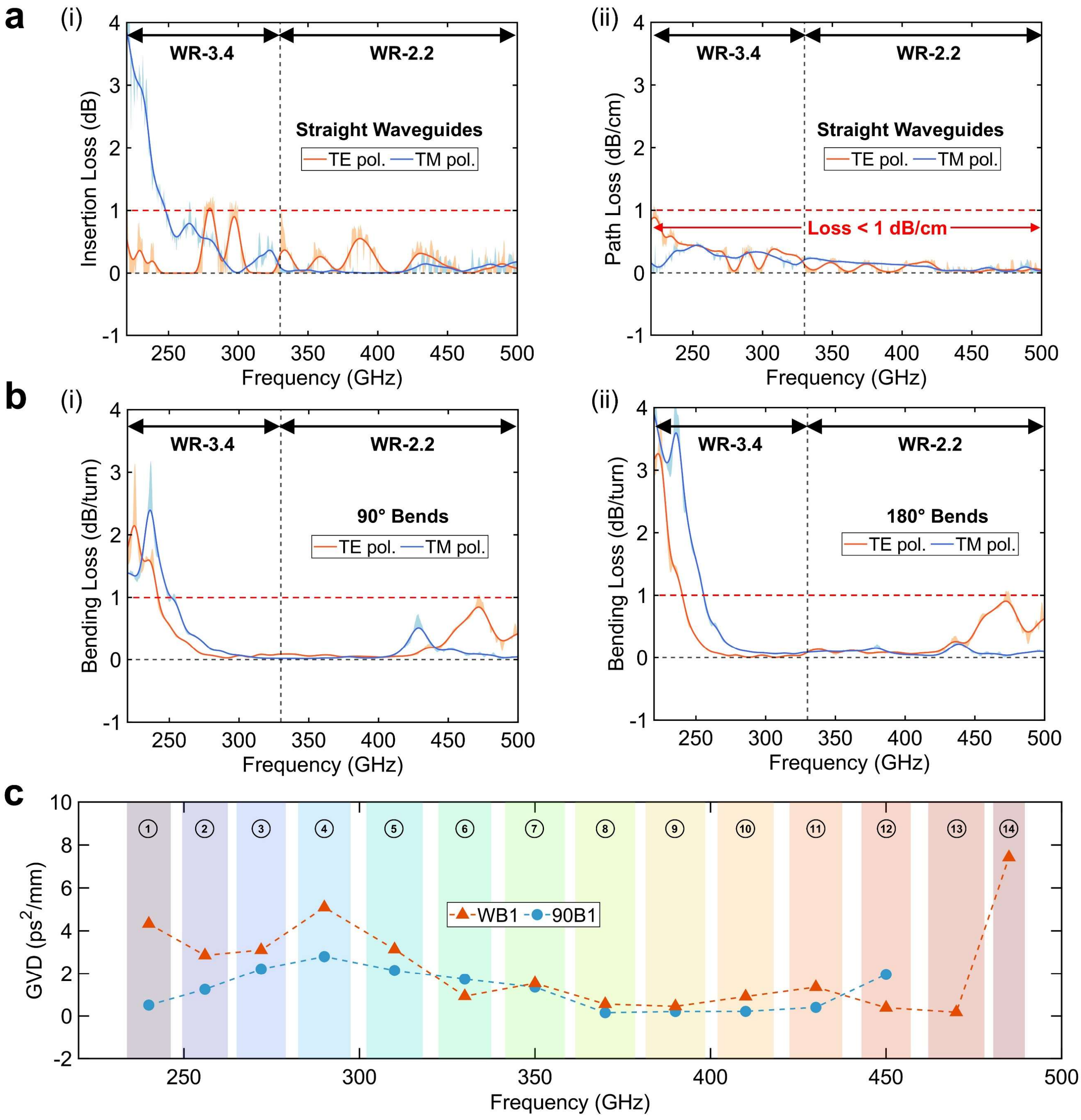

**Fig. 3 | Characterizations of device losses and group velocity dispersion (GVD) in all-silicon straight and bent waveguides. a**, Calculated (i) insertion loss and (ii) path loss for both TE and TM polarizations of the straight interconnects, obtained by linear fitting of the measured transmission data for WB1, WB2, and WB3 samples. The red dashed lines in (**a**-i) and (**a**-ii) indicate the 1 dB insertion loss and 1 dB/cm path loss levels, respectively. For the path loss, red arrows highlight the frequency regions where the loss remains consistently below 1 dB/cm, spanning the entire WR-3.4 and WR-2.2 bands. **b**, Calculated bending loss for (i) 90° bends and (ii) 180° bends for both TE and TM polarizations. The bending loss is extracted by subtracting the insertion and path losses from the measured transmission of the 90° (90B1, 90B2, and 90B3), and 180° (180B1, 180B2, and 180B3) bent waveguides, followed by linear fitting of the resulting transmission data. The red dashed lines in (**b**-i) and (**b**-ii) indicate a bending loss of 1dB per

turn. The orange and blue shaded regions in (**a**-i–ii) and (**b**-i–ii) represent the variation of the losses relative to the smoothed loss values. **c**, Calculated absolute GVD for the WB1 and 90B1 samples, obtained by linear fitting of the measured group-delay data over the individual channel bandwidths. GVD is evaluated for fourteen channels in the WB1 sample and twelve channels in the 90B1 sample, with individual channels indicated by shaded regions and labelled by channel number.

We quantify the loss mechanisms and dispersion characteristics of the THz silicon waveguide platform (see Supplementary Information S4 for details of the loss-characterization methodology). By linearly fitting the measured transmission ($S_{21}$) data from straight waveguides of different transmission link lengths (WB1, WB2, and WB3), we independently extract the insertion and path losses for both TE and TM polarizations, as shown in Fig. 3a(i) and (ii). Furthermore, the average insertion, path, and bending losses are calculated from the extracted data over the 220-500 GHz range. The insertion loss remains below 1 dB over majority of the operating band. Coupling of the THz energy between the hollow metallic waveguides and the integrated tapered couplers contributes average insertion losses of 0.21 dB and 0.48 dB for TE- and TM-polarized THz waves, respectively. The measured path losses remain well below 1 dB/cm across the WR-3.4 and WR-2.2 bands, with average values of 0.22 dB/cm and 0.19 dB/cm for TE and TM polarizations, respectively. Minimum path losses of 0.016 dB/cm at 391.06 GHz (TE) and 0.024 dB/cm at 475.94 GHz (TM) further underscores the inherently low-loss nature of the HR-Si platform. This analysis is extended to bent waveguides, where bending loss for 90° and 180° turns are isolated by subtracting the insertion and path loss contributions from the measured transmission (Fig. 3b(i), (ii)). For both polarizations, the bending loss remains well below 1 dB per turn across the operational 3-dB bandwidth, confirming that tight subwavelength bends impose only a modest penalty. Average bending losses are 0.36 dB and 0.37 dB per turn for TE and TM waves in 90° bends, respectively, and 0.39 dB and 0.56 dB per turn in 180° bends. Minimum bending losses of 0.0033-0.0329 dB per turn are observed for both configurations and polarizations, as also shown in Fig. 3b(i, ii), highlighting the negligible bending loss introduced by adiabatic bending. The orange- and blue-shaded regions in the loss plots indicate the variation relative to the smoothed loss trends. In addition to the loss analysis, we assess dispersion by extracting the absolute group-velocity dispersion (GVD) from linear fits to the measured frequency-domain group-delay data for TE-polarized transmission in the straight waveguide (WB1) and the 90° bent waveguide (90B1). Low GVD is critical for high-speed data communication, as it minimizes pulse broadening and intersymbol interference (ISI), thereby preserving signal integrity across THz channels. The GVD analysis is performed for the non-simultaneously excited channels used in the on-chip THz communication experiments (Fig. 3c; see Supplementary Information S5). Specifically, we extract the group-delay response for each channel in both WB1 and 90B1 and apply linear fitting to determine the per-channel GVD values. WB1 and 90B1 exhibit GVD values ranging from 0 to 8 $ps^2/mm$, with WB1 approaching the upper limit at the lower and upper edges of the operating bandwidth. In contrast, 90B1 shows smaller variations, with most channels exhibiting values below 4 $ps^2/mm$ owing to the flatter group-delay response. Most channels exhibit exceptionally low absolute GVD, with minimum values of 0.18 $ps^2/mm$ (WB1) and 0.15 $ps^2/mm$ (90B1). Such low dispersion minimizes temporal pulse broadening during propagation, thereby preserving symbol integrity, reducing ISI, and enabling complex modulation formats in the THz communication experiments. The group-delay responses measured for TM-polarized transmission in WB1 and 90B1 are similarly flat. We therefore expect comparably low absolute GVD values across multiple channels for on-chip THz communication using TM-polarized waves, in addition to the TE polarization employed in the present configuration.

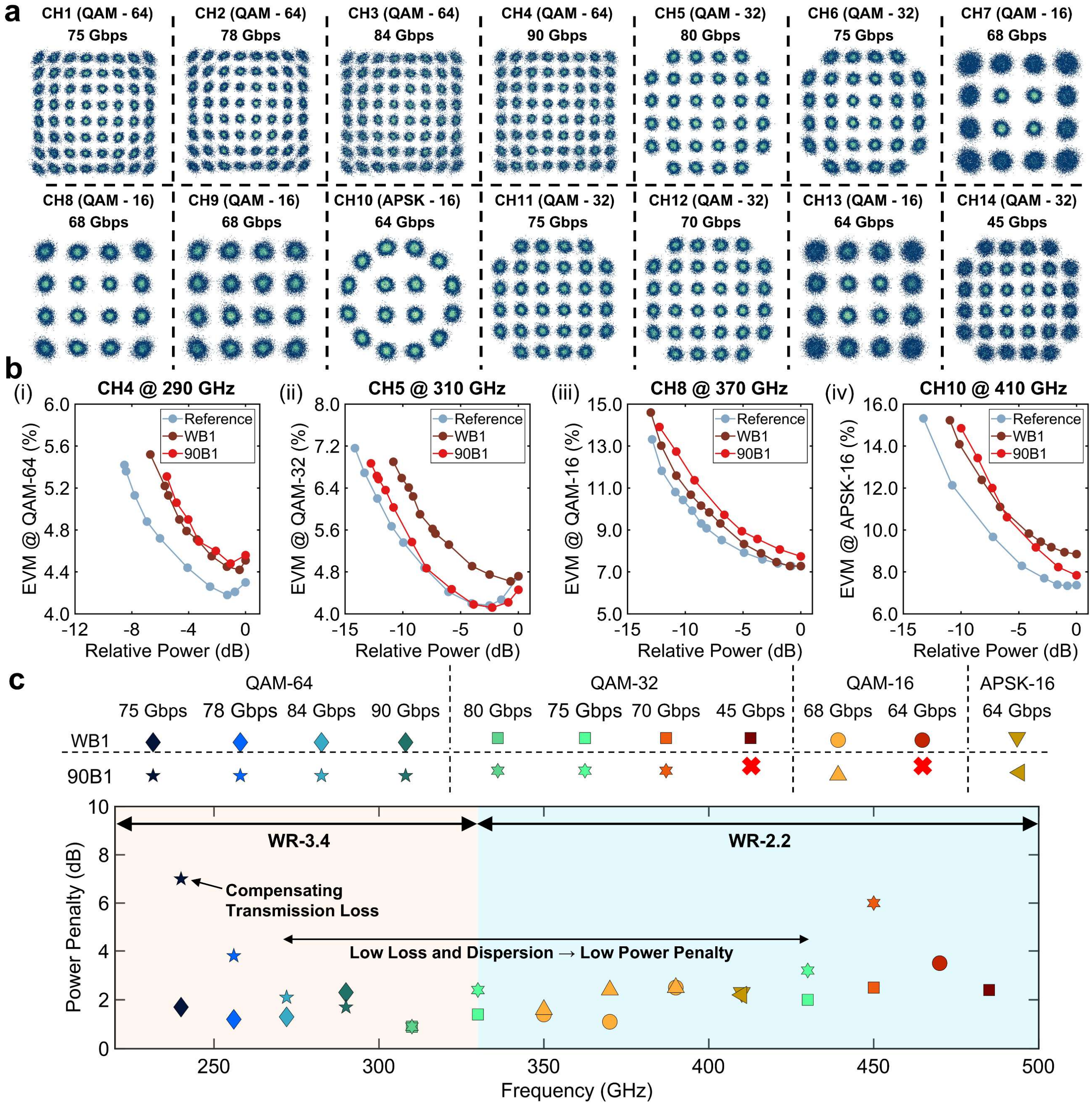

**Fig. 4 | On-chip terabit-per-second THz communication using complex modulation formats including QAM-16, QAM-32, QAM-64, and APSK-16. a**, IQ constellation diagrams for the straight waveguide (WB1) DUT, showing the successful transmission of fourteen channels across the WR-3.4 and WR-2.2 bands. The well-separated symbol clusters for the different modulation formats highlight the extremely low-distortion characteristics of the designed waveguide. The corresponding data transfer rates and modulation formats for each channel are also indicated. **b**, Measured EVM versus relative power for channels (i) 4 (QAM-64), (ii) 5 (QAM-32), (iii) 8 (QAM-16), and (iv) 10 (APSK-16), at channel centre frequencies of 290, 310, 370, and 410 GHz, respectively, for the back-to-back (B2B) reference, WB1, and 90B1 samples. These channels are selected as they exhibit the highest data transfer rates among the four different complex modulation formats. **c**, Power penalty versus channel frequency for individual channels employing complex modulation formats in the WB1 and 90B1 devices. Identical

data rates are achieved across corresponding channels in WB1 and 90B1 by compensating for the higher transmission loss in 90B1, resulting in increased power penalties below 275 GHz in the WR-3.4 band and above 425 GHz in the WR-2.2 band. The low-loss and low-dispersion frequency regions correspond to reduced power penalties. The red crosses mark the channels (13 and 14) that are not used in 90B1 for the on-chip THz data communication, in-line with the 90B1 transmission curve (Fig. 2c(iii)).

Using a photonics-based transmitter (Tx) and an electronics-based receiver (Rx) THz communication setup (see Supplementary Information S1), we characterize the straight waveguide (WB1) and its 90° bent counterpart (90B1) to directly assess the impact of 3-dB bandwidth and GVD on the achievable data rates. We also measure a back-to-back (B2B) reference, in which we directly connect the hollow metallic waveguides without inserting the device under test (DUT), to assess the testing system performance for all tested channels. All data transmission experiments are conducted using TE-polarized THz waves. Given the similarly broad bandwidth and low GVD for TM polarization, we expect comparable data rates under TM operation. In Fig. 4, we demonstrate terabit-per-second on-chip THz data transmission using high-order modulation formats, including quadrature amplitude modulation (QAM) and amplitude-phase-shift keying (APSK). Fourteen non-simultaneously excited channels spanning the WR-3.4 and WR-2.2 bands (220-500 GHz) are transmitted using QAM-16, QAM-32, QAM-64, and APSK-16, as verified by the IQ constellation diagrams measured for WB1 (Fig. 4a). The well-defined constellation points show minimal waveform distortion and validate the combined low-loss and low-dispersion performance of the waveguide platform (see Supplementary Information S6 for B2B and 90B1 IQ constellations). We achieve a maximum single-channel data rate of 90 Gbps on channel 4 using QAM-64, and peak data rates of 80, 68, and 64 Gbps on channels 5, 8, and 10, respectively, using QAM-32, QAM-16, and APSK-16. System performance is further quantified by measuring the error vector magnitude (EVM) as a function of relative power. Figures 4b(i–iv) present EVM versus relative power for channels 4, 5, 8, and 10, centred at 290, 310, 370, and 410 GHz, respectively, and employing QAM-64, QAM-32, QAM-16, and APSK-16 (see Supplementary Information S7 for all channel EVMs). Compared with the B2B reference, WB1 and 90B1 introduce only modest EVM degradation, confirming that propagation and bending do not cause significant signal impairment, even at carrier frequencies above 400 GHz. Given that the transmission dip in 90B1 limits the usable bandwidth, only twelve channels are excited in this device, compared with fourteen in WB1 and the B2B reference. Finally, we extract the power penalties from the EVM versus relative power characteristics for all THz channels. Power-penalty analysis quantifies the additional received power required to maintain a target EVM in the presence of impairments from the waveguide under test. The extracted penalties are summarized in Fig. 4c. For identical data rates, 90B1 requires higher relative received power to compensate for additional transmission losses, leading to increased power penalties at the band edges (<275 GHz and >425 GHz). In contrast, the low-loss, low-dispersion frequency regions incur minimal penalties (<4 dB). These results collectively demonstrate the robustness of the platform to high-order modulation and complex waveguide routing.

In conclusion, we demonstrate on-chip terabit-scale THz data link using dispersion-engineered, effective-medium-supported, unclad silicon waveguide interconnects fabricated on HR-Si. The all-silicon platform supports multi-band operation across two adjacent THz bands (WR-3.4 and WR-2.2), spanning 220-500 GHz, with low path loss, low bending loss, and low GVD for both TE and TM polarizations. By systematically disentangling insertion, path, and bending losses, we show that the waveguides exhibit average path losses below 1 dB/cm and average bending losses below 0.6 dB per turn over the full operating band. Using a combined photonic transmitter (Tx) and electronic receiver (Rx) architecture, aggregate data rates of 1.004 Tbps in straight waveguides and 0.895 Tbps in 90° bent waveguides are achieved using TE-polarized THz waves, enabled by GVD as low as 0.15 $ps^2$/mm. Our work establishes dispersion-engineered, effective-medium-supported, unclad silicon waveguides as a potentially scalable platform for multi-THz-band, multi-channel, low-loss THz interconnects, and points toward their application in future on-chip, chip-to-chip, and hybrid electronic-photonic THz systems with enhanced spectral efficiency.

## Methods

### Sample Fabrication

All-silicon interconnect chips were fabricated on 6-inch high-resistivity (>20 kΩ.cm) silicon (HR-Si) wafers with a thickness of 500 μm. Following standard wafer cleaning, a 2.4 μm-thick silicon dioxide ($SiO_2$) layer was deposited as an etch-protection mask. Conventional ultraviolet (UV) photolithography, followed by reactive ion etching, was used to define the circular and semi-circular etch holes of the effective-medium cladding, the unclad regions of the devices, and the device separation trenches on the $SiO_2$ layer. The remaining pattern on the $SiO_2$ layer works serves as a protective mask, after which the silicon wafer was etched using deep reactive ion etching (DRIE, Oxford Estrelas) to a depth of 370 μm, ensuring uniform and smooth sidewalls. Finally, any residual $SiO_2$ was subsequently removed, and the silicon wafers were backgrinded to a final thickness of 220 μm.

### Experimental Setup

Vector Network Analyzer (VNA) measurements were performed using a Rohde & Schwarz ZNA26 network analyzer, which generates microwave signals in the 10 MHz to 26.5 GHz range. These signals were upconverted to the WR-3.4 (220-330 GHz) and WR-2.2 (330-500 GHz) frequency bands using frequency extenders in WR-3.4 (ZC330) and WM-570 (WR-2.2) Z500 from Rohde & Schwarz. The generated THz signals were coupled into the all-silicon interconnect chip via WR-3.4 or WR-2.2 metallic waveguides through the input taper. The transmitted THz signals were collected using the corresponding output waveguides and directed into the VNA port 2, where they were down-converted to an intermediate frequency (IF) using a local oscillator (LO) signal generated by the network analyzer. Prior to measurement, the system was calibrated using the thru-reflect-match (TRM) waveguide calibration procedure in accordance with WR-3.4 and WR-2.2 standards. This calibration compensates for system non-idealities and ensures accurate measurement of the chip response. After calibration, the transmission was normalized to 1 (0 dB) with a direct probe to probe connection, and the group delay, representing the signal transit time through the device under test (DUT), was set to 0 ns. The WR-3.4 or WR-2.2 waveguides used for coupling were then connected in a back-to-back configuration, and the corresponding transmission and group delay were measured. The chip transmission and group delay were obtained by subtracting the reference responses from the overall system measurements. Details of the experimental setup for the on-chip terabit-per-second THz communication experiment are provided in Section S1 of the Supplementary Information.

## AUTHOR INFORMATION

### Corresponding Author

*Ranjan Singh (rsingh3@nd.edu).

### Author Contributions

B.C. and R.S. conceived the idea; B.C. performed numerical analysis with the help of W.W. and N.N.; B.C., W.W., N.N., and T.C.T. performed the initial VNA measurements; G.D. performed the final reported VNA measurements. P.S., H.N.K., and G.D. performed the on-chip THz communication experiments; B.C. wrote the initial draft of the manuscript; W.W., N.N., T.C.T., P.S., H.N.K., G.D., and R.S. revised the manuscript; R.S. led the overall project.

### Acknowledgments

B.C., W.W., N.N., T.C.T., and R.S. acknowledge the support from the National Research Foundation (NRF) Singapore, Grant No: NRF-MSG-2023-0002. B.C., W.W., N.N., T.C.T., and R.S. thank Ling Ling Ngo from Nanyang NanoFabrication Centre (N2FC) for the help with sample fabrication, and Xavier Low and Anna Koh from Disco Corporation for the help with wafer backgrinding process. P.S., H.N.K., and G.D. state that the characterization testbeds are supported by France 2030 programs, PEPR (Programmes et Equipments pour la Recherche) and CPER Wavetech. The PEPR is operated by the Agence Nationale de la Recherche (ANR), under the grants ANR-22-PEEL-0006 (FUNTERA, PEPR 'Electronics') and ANR-22-PEFT-0006) NF-SYSTERA, PEPR 5G and beyond – Future Networks). We also acknowledge the Renatech network as well as the Nanofutur program, grant 21-ESRE-0012. The Contrat de Plan Etat-Region (CPER) Wavetech is supported by the Ministry of Higher Education and Research, the Hauts-de-France Region council, the Lille European Metropolis (MEL), the Institute of Physics of the French Regional Centre for Scientific Research (CNRS), and the European Regional Development Fund (ERDF). The authors acknowledge support from CDP C2EMPI, the French State under the France-2030 program, the University of Lille, the Initiative of Excellence of the University of Lille, the European Metropolis of Lille for their funding and support of the R-CDP-24-004-C2EMPI project.

**Competing interests**

The authors declare no competing interests.